\documentclass[a4paper]{article}

\usepackage{icrc2013}

\title{Large Size Telescope camera support structures for the Cherenkov Telescope Array}

\shorttitle{LST camera support structures for the Cherenkov Telescope Array}

\authors{
G. Deleglise$^{1}$, N. Geffroy$^{1}$, G. Lamanna$^{1}$ 
for the Cherenkov Telescope Array Consortium.
}

\afiliations{
$^1$ LAPP, Universit\'{e} de Savoie, CNRS/IN2P3, F-74941 Annecy-le-Vieux, France.
}

\email{Giovanni.Lamanna@lapp.in2p3.fr}

\abstract{The design of the camera support structures for the Cherenkov Telescope Array (CTA) Large Size Telescopes (LSTs) is based on an elliptical arch geometry reinforced along its orthogonal projection by two symmetric sets of stabilizing ropes. The main requirements in terms of minimal camera displacement, minimal weight, minimal shadowing on the telescope mirror, maximal strength of the structures and fast dynamical stabilization have led to the application of Carbon Fibre Plastic Reinforced (CFPR) technologies. This work presents the design, static and dynamic performance of the telescope fulfilling critical specifications for the major scientific objectives of the CTA LST, e.g. Gamma Ray Burst detection.}


\begin{document}
\maketitle


\section{The CTA project}
The Cherenkov Telescope Array (CTA)~\cite{bib:cta} is the next generation system of Very High Energy (VHE) gamma-ray Imaging Atmospheric Cherenkov Telescopes (IACT) and the first open-access astronomical observatory operating in the energy range from a few tens of GeVs to $\sim$100 TeV.

The CTA project, currently in its pre-construction phase, is led by a worldwide international consortium aiming to deploy about 100 Cherenkov telescopes and in two sites in the two hemispheres for a complete sky coverage. CTA consists of three types of telescopes: the Large Size Telescopes (LSTs) sensitive to the lower energies, the Mid Size Telescopes to guarantee the high sensitivity in the range 100 GeV to a few tens of TeVs, and Small
Size Telescopes to explore the highest energies.

\begin{figure*}[!]
  \centering
\includegraphics[width=0.53\textwidth]{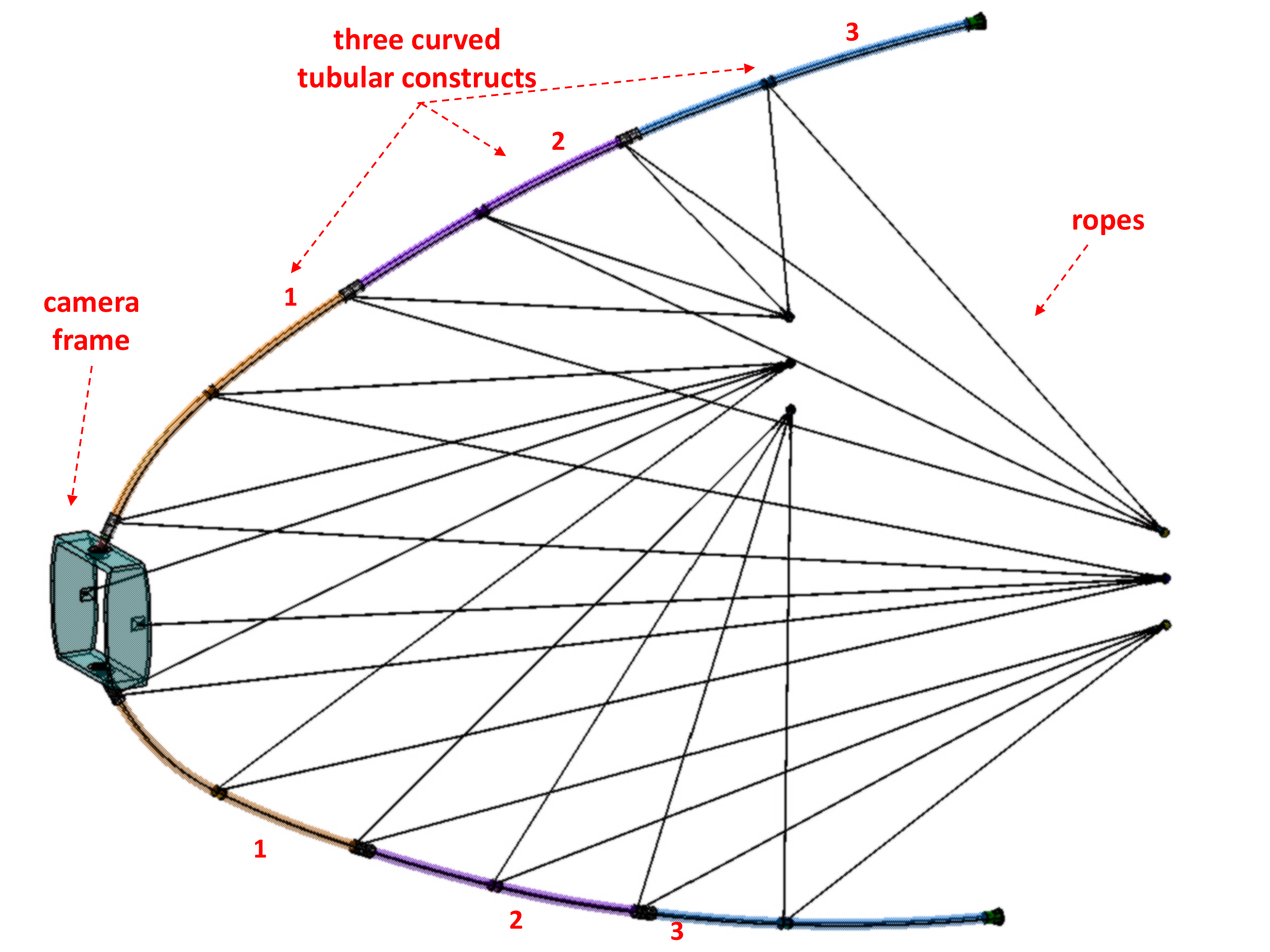}
\includegraphics[angle=+90,width=0.3\textwidth]{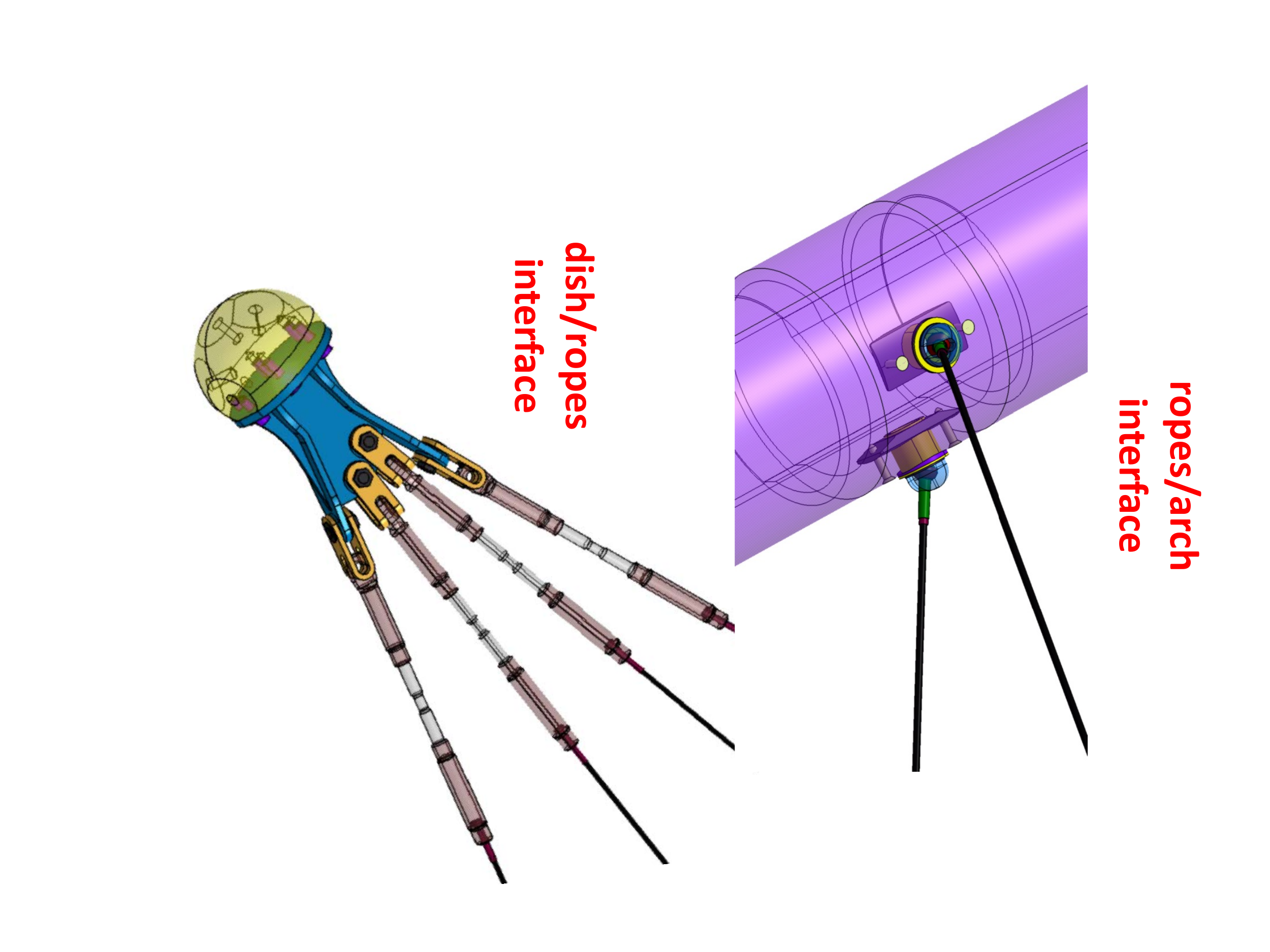}
  \caption{Design scheme of the CSS keeping the LST camera at a focal distance of 28 m and for 23 m dish diameter.}
  \label{arch}
 \end{figure*}

\subsection{The Large Size Telescopes}
The main observation domain of the LSTs will be in the energy range between the threshold of approximately 20 GeV and about 200 GeV. This energy domain is particularly rich and in certain aspects unique in important physics questions such as: the upper end of the gamma-ray emission of pulsars; AGN spectra around the cut-off set by the interaction of high energy gamma-rays with the EBL and at different redshifts; measurements of the temporal and spectral distributions of the GRBs at different redshifts; indirect search for dark matter; and testing possible Lorentz invariance violation. The CTA physics cases set a number of challenging structural constraints to the LST design~\cite{bib:oscar} among which are: a long focal distance to optimise the optical performance of the telescope; rapid positioning onto GRBs and a high level of dynamical mechanical stability of the structures supporting the camera to minimise the camera misalignment and optimise the angular resolution. Extended Monte Carlo simulation studies have been conducted in order to explore and decide on the basic design parameters for the telescope and these have been fixed at a focal length F=28 m, field of view FoV=4.5$^{\circ}$, pixel size fixed to 0.1$^{\circ}$, and a dish diameter D=23 m for a ratio F/D=1.2. A specification of 20 s maximum delay to re-point the LSTs in the direction of a GRB alert at 180$^{\circ}$ angular distance from the previous pointing position is one more severe constraint on the telescope system design.

\section{The LST camera support structure}

The LST Camera Support Structure (CSS) is designed to respond to critical requirements aimed at guaranteeing the best optical properties of the telescope, and to provide a high level of mechanical stability (static and dynamic), while minimising both the weight of the masts and the shadow projected onto the mirror due to their geometry. These last constraints have implied the adoption of composite materials, namely Carbon Fibre Plastic Reinforced (CFPR) structures. Indeed, CFPR materials are well known to provide a very high performance-to-mass ratio.
The CSS is also called an $arch$ making reference to its planar pseudo elliptic geometry (see figure \ref{arch}). The arch design consists of three CFPR curved tubular constructs of circular section, with diameter 310 mm and thickness 14 mm, for each of its two symmetric arms. A CFPR camera frame has an internal square space of 3140$\times$3140 mm, designed for the sliding and fixing of the octagonal camera and square blinds. 26 CFPR ropes aligned along its transverse projection and of lengths between 17 m and 28 m are applied to stiffen the structure: although these ropes aim to preserve the out-of-plane stiffness, they also contribute to the in-plane stability of the structure (camera sagging as a function of elevation) and therefore of the overall dynamics of the arch. The pre-tension sets are (for a 0$^{\circ}$ telescope elevation) in the range between 900 kg and 2 tons,
depending on the group of ropes. The devices fixing the ropes to the dish and the arch are designed by naval designers (and commonly used to link masts to ropes).

In order to get rid of thermal expansion difference issues all junction elements between CSS components are made of CFPR. Stainless steel parts are used only in the arch/dish interfaces for higher linking efficiency and production process purposes.

The manufacturing process which we propose to use for the arch components and the camera frame is known as $pre$-$preg$ technology, used for state-of-the-art structures such as racing boats and cars, new generation aircraft etc.

The main advantages of this technique for the LST purposes are:

- the possibility of achieving any fibre orientation;

- the possibility of reinforcing the structures on well localised critical parts;

- a lower resin-to-fibre ratio;

- a better compaction of the layers enabled by the curing process (autoclave);

- a higher glass transition temperature guaranteeing stable performance at high environmental temperatures over a long time period.
  \begin{figure}[!b]
  \centering
  \includegraphics[height=0.4\textwidth,width=0.48\textwidth]{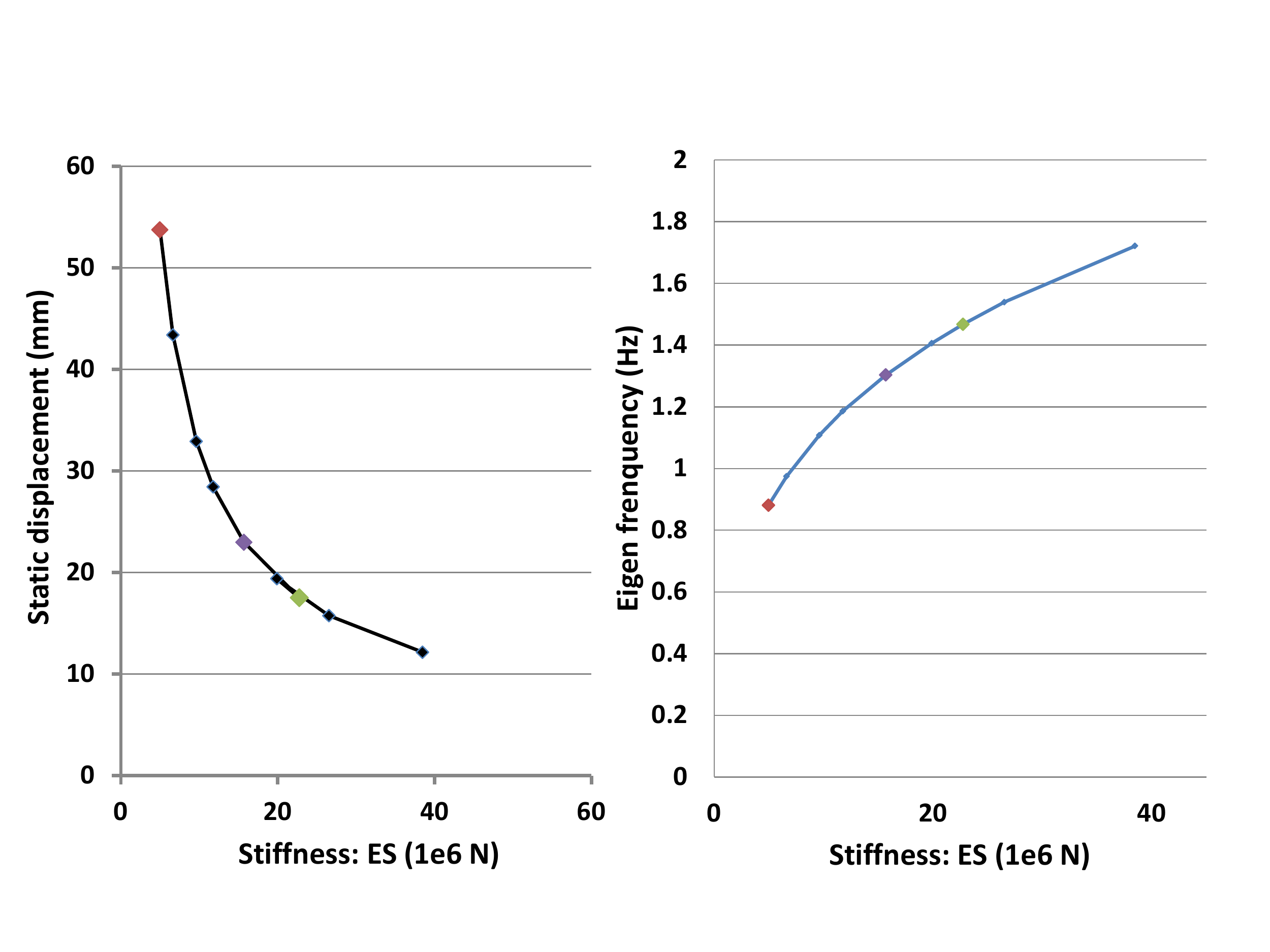}
  \caption{CFPR rope stiffness computed as the product between  Young's modulus $E$ and the rope section $S$ affecting the static camera displacements and the first telescope eigenfrequency.}
  \label{wide_fig}
 \end{figure}
Moreover existing finite element programs enable a precise modelling of the $pre$-$preg$ technology thanks to the precise knowledge of fibre orientations along the entire structure. Consequently the mechanical design and simulations have provided highly detailed, reliable and reproducible results. For instance it has permitted the control, calibration and modification of the tubes thickness, which is kept constant all along the arch components, except when more stiffness is demanded. 
In contrast, the CFPR ropes are produced using a $filament$ $winding$ process in order to get the following optimal performances: - all CF fibres are along 0$^{\circ}$ (axial) orientation; - metallic end-fittings are included during the process without cutting fibres (no gluing, no fastening system required). 
The effect of rope stiffness computed as Young's modulus $E$ (in the range of 150-300 GPa) $\times$ the rope section $S$ (in the range of 3-to-15 mm diameter) were studied versus some critical figures of merit, e.g. camera displacement and first eigen-frequencies of the telescope, as shown in figure \ref{wide_fig}. Ropes’ of 21$\times$10$^{6}$ N (for $E$ = 270 GPa; $S$ = 78.5 mm$^{2}$ for 10 mm diameter) stiffness were selected for the LST-CSS. 

The resulting mechanical design of the CSS counts for a total mass of 2513 kg and has been object of detailed finite element analysis, the results of which are presented in the next section.

\subsection{Finite element and performance analysis}

The CSS finite element model has been realised using shell elements and considering the arch presented earlier. The overall structure is meshed with shell elements, and the composite material is introduced through a stacking sequence of orthotropic plies.
Local reinforcement areas have been explicitly modelled by varying this stacking sequence, especially for rope anchoring and composite tube junctions. The camera frame, built with sandwich technology, has been simulated with the same level of detail. In this case, a set of special elements ensure the fixation of the 2.5 tons camera to rail attachment positions~\cite{bib:oscar}.

\begin{figure}[!t]
  \centering
   \includegraphics[width=0.38\textwidth]{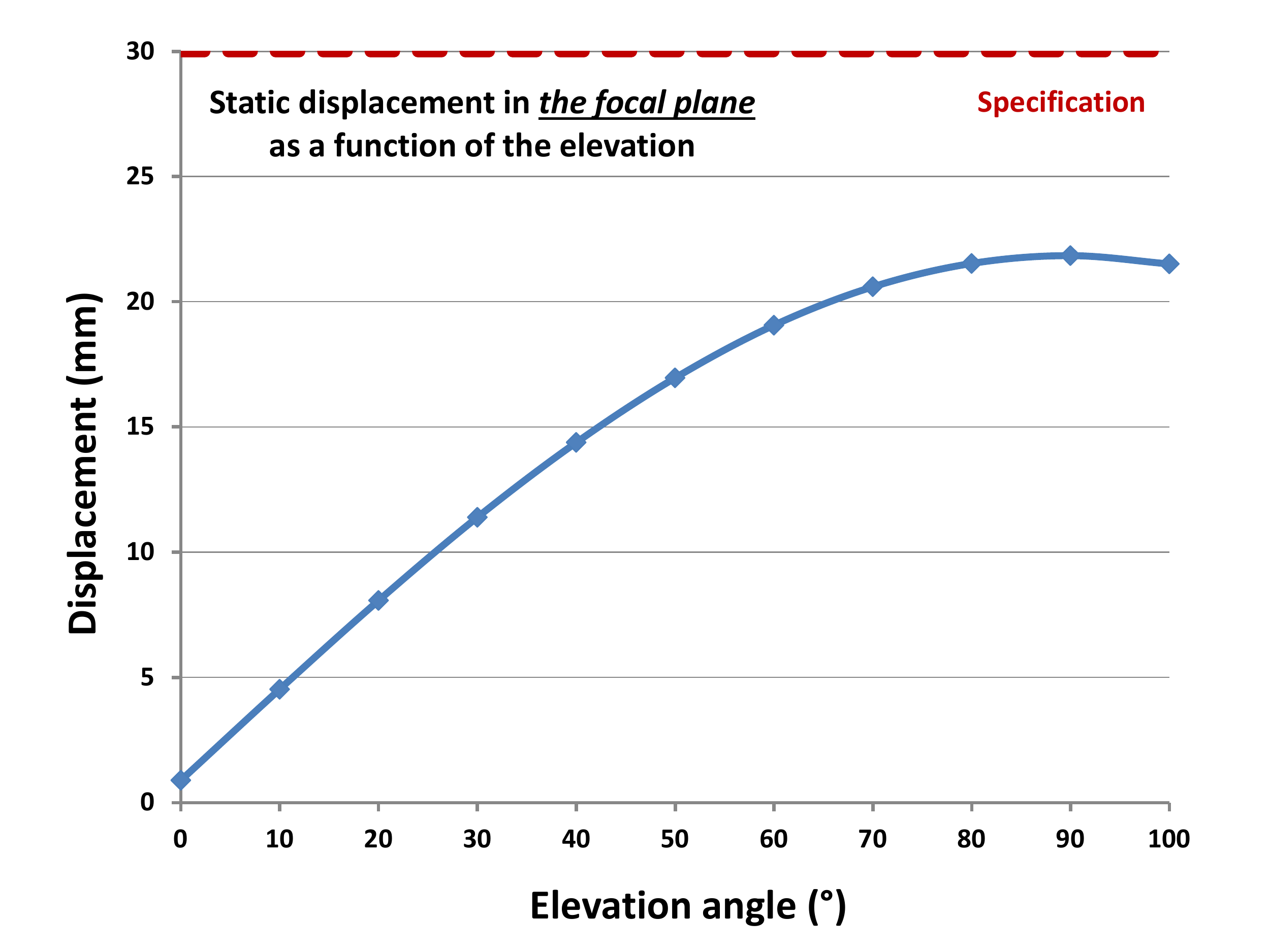}
   \includegraphics[width=0.38\textwidth]{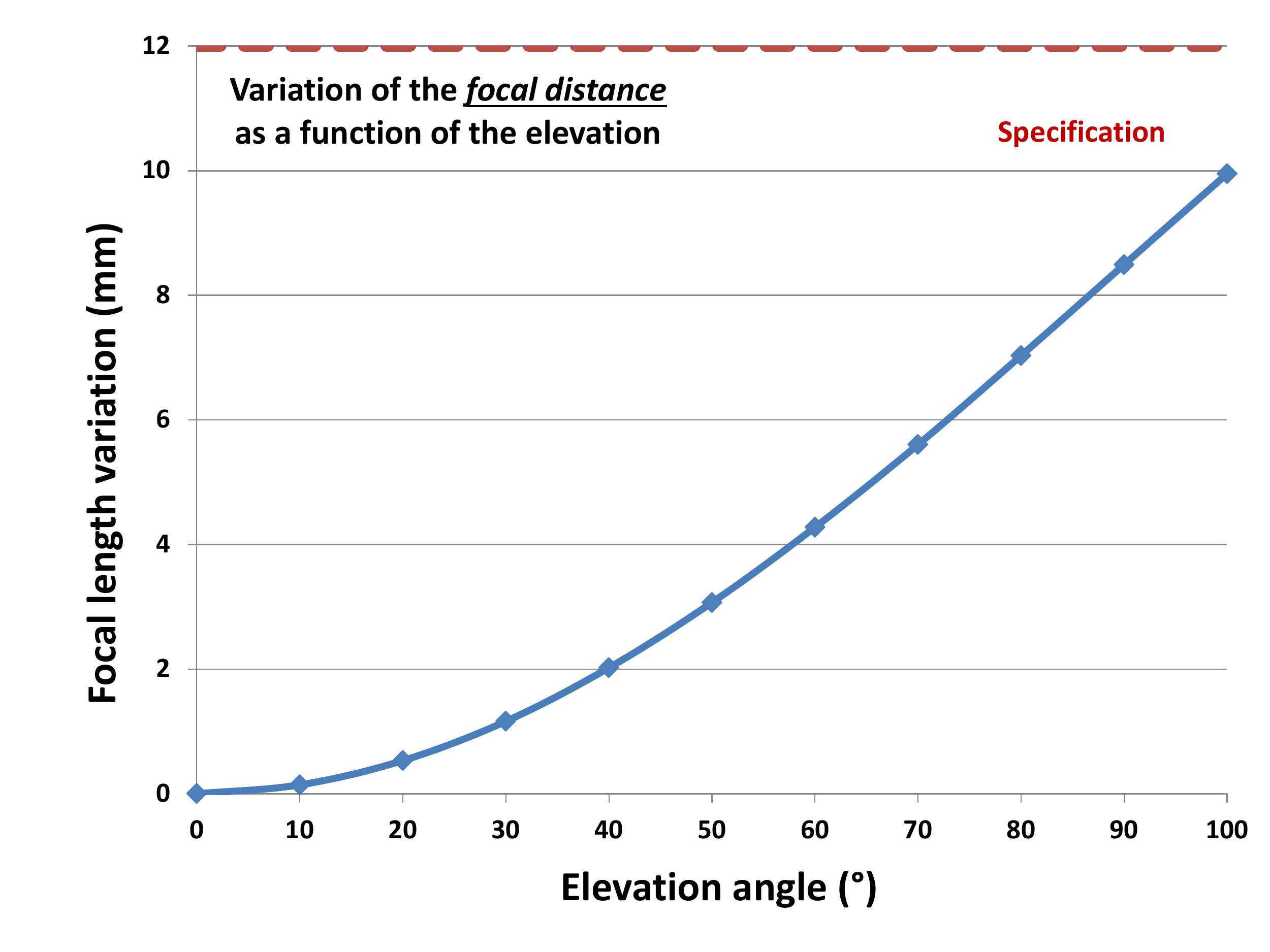}
   \includegraphics[width=0.38\textwidth]{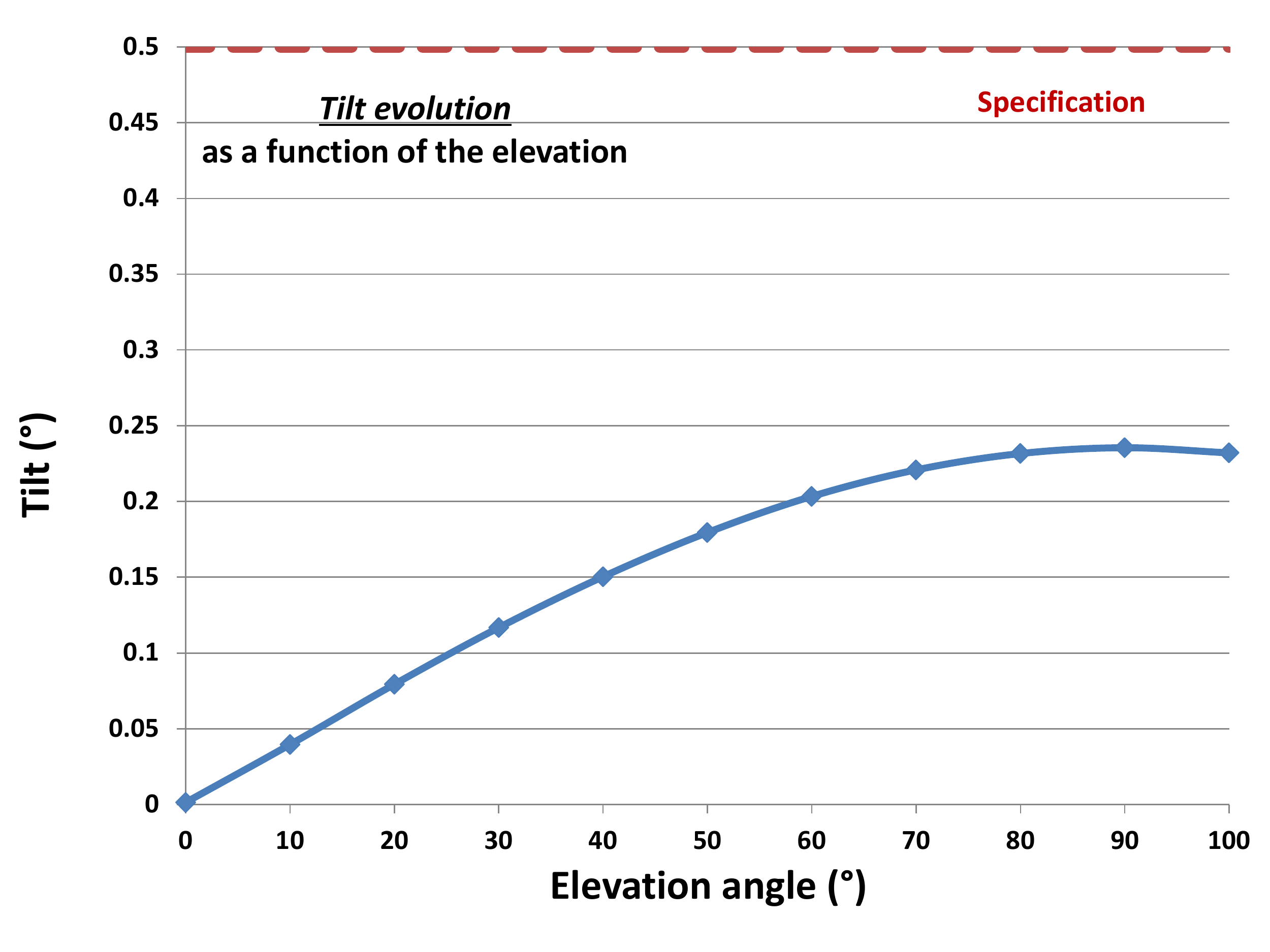}
  \caption{Camera static displacements.}
\label{static}
 \end{figure}

In addition, the behaviour of the CSS has been investigated in detail taking into
account the equivalent stiffness for each interface point (two for the mast interface and six for the rope interfaces). Interface point stiffnesses were computed by considering only the sub-structures (mount \& dish), without gravity, under several unitary load cases. Then the equivalent coupled stiffness matrices associated with each interface point were derived for a 0$^{\circ}$ telescope elevation, under the assumption that mount and dish stiffness are independent of the elevation. The defined equivalent values of stiffness, computed from inputs provided by other partners of the LST project, were used both in static and dynamic calculations.

The CCS mechanical behaviour was studied under different static, dynamic, and thermal expansion load cases. Furthermore, critical situations affecting the data-taking phases of the telescopes were also simulated, namely the telescope behaviour under the action of turbulent wind, after fast repositioning, e.g. following-up a received GRB alert, and fast emergency braking. 
In the following just a few outstanding results will be illustrated aiming to provide an overview of the CSS performance achieved with the current design. All these results have to be considered as preliminary and will be confirmed in the near future by a prototype testing campaign.

\subsubsection{Static load cases}

The first case study is the inspection of camera static displacements and tilt after considering only gravitational loads on the CSS as a function of telescope elevation. Three different camera static displacements are investigated which could imply some level of optical misalignment of the camera, potentially affecting the angular resolution required to be achieved by the LST:

1) Camera shift along the focal plane and therefore the consequent displacement of the point of intersection of the optical axis from the geometrical centre of the camera. The design specification corresponds to a maximum acceptable displacement of 30 mm. The current CSS design fulfils this by keeping the displacement below 17 mm for elevation between 0 and 45$^{\circ}$ and less than 22 mm for elevation of 90$^{\circ}$.
 
2) The movement of the camera along the focal axis. This is limited to 3 mm for elevations up to 45$^{\circ}$, growing to 10 mm for elevations up to 100$^{\circ}$, which is well within the specified maximum 12 mm movement.

3) Camera tilt around the axis orthogonal to the optical axis. The outstanding stability of the current design allows the arch to limit such a camera displacement at less than 0.25$^{\circ}$ well below the specification (0.5$^{\circ}$).

All the above results are illustrated in figure \ref{static}.

\subsubsection{Dynamic load cases}

Several load cases have been considered in realizing the sizing of the structure. They may be grouped into three main scenarios: 

- Data taking : displacement criteria (sag, tilt, dynamic displacement due to turbulent wind). 

- Gamma ray burst repositioning : integrity during movement \& damping within 10 s after end of movement. 

- Security and emergency stop : integrity of structure.\\
More precisely seventeen dynamic calculations combining the effects of wind (in three different directions) and different telescope acceleration and braking phases were performed together with eight more computations of structural thermal expansions and six seismic load cases (static equivalent loading).   

In order to show that the structure described above matches the requirements, dynamic results of two specific load cases for GRB repositioning are illustrated below.

 \begin{figure}[!t]
  \centering
  \includegraphics[width=0.33\textwidth]{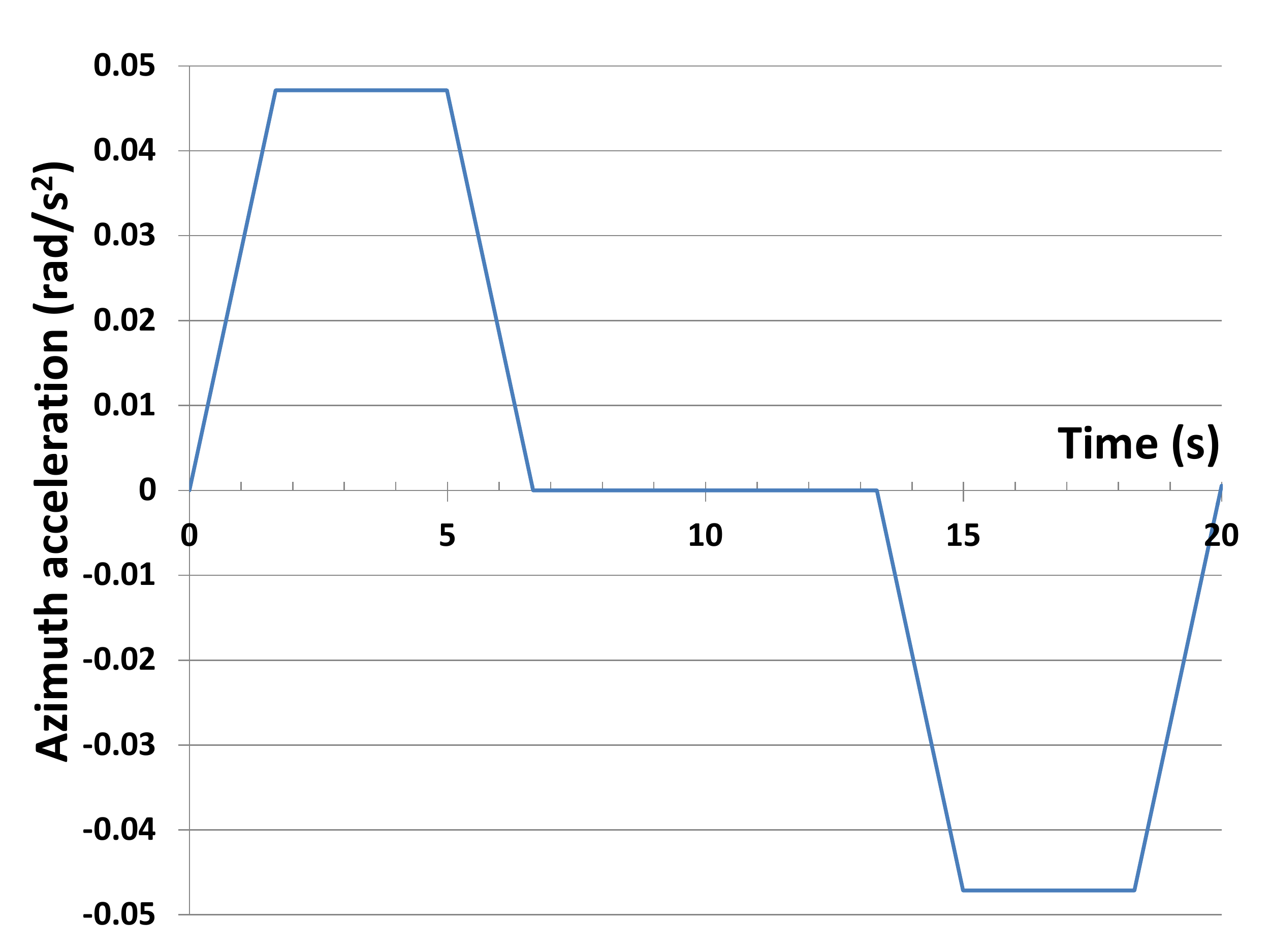}
  \caption{Angular azimuth acceleration evolution of the LST moving in 20 s onto a GRB at 180$^{\circ}$ distance apart.}
\label{ino13}  
 \end{figure}

\begin{figure}[!b]
  \centering
  \includegraphics[width=0.4\textwidth]{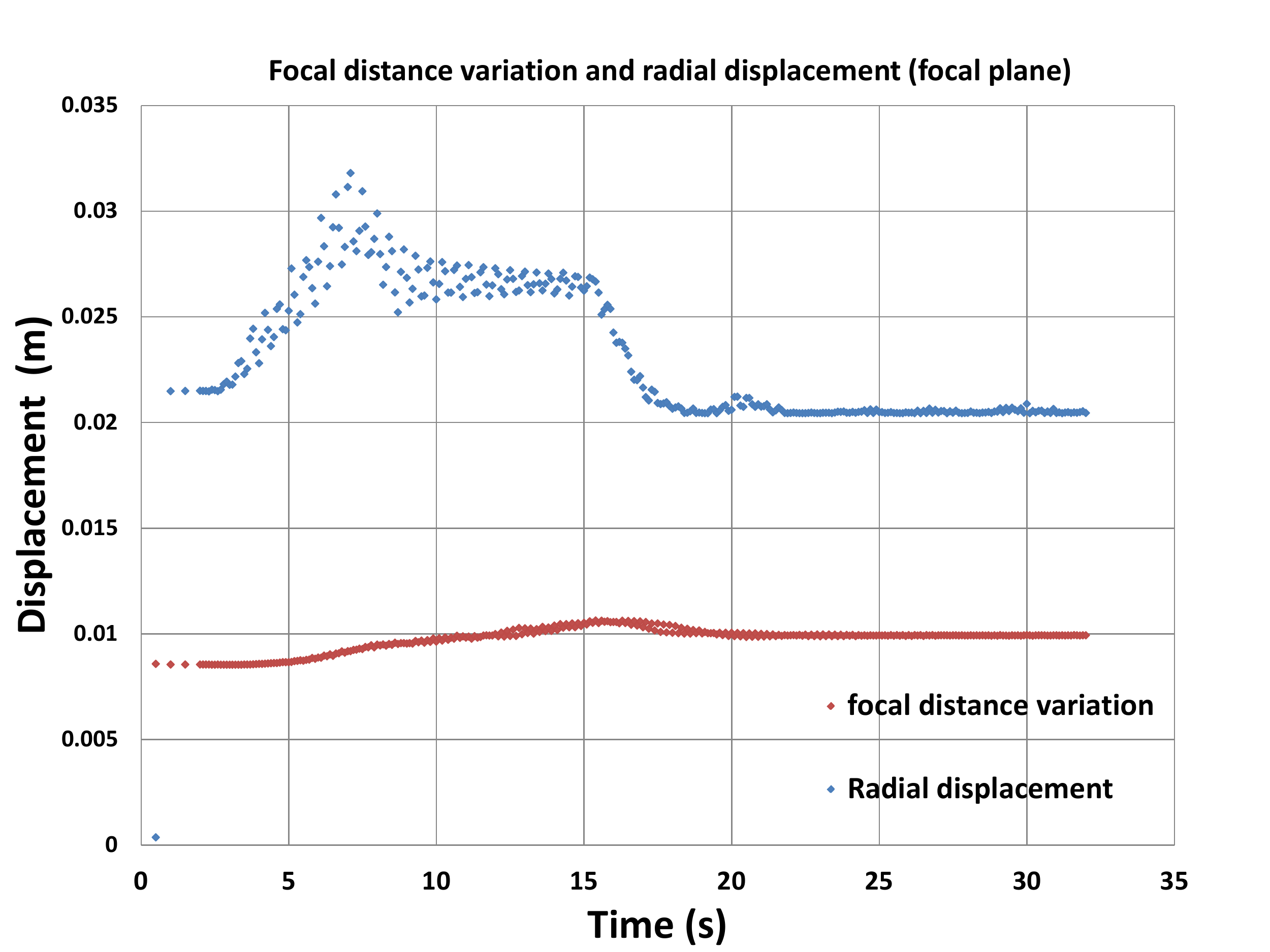}
  \includegraphics[width=0.4\textwidth]{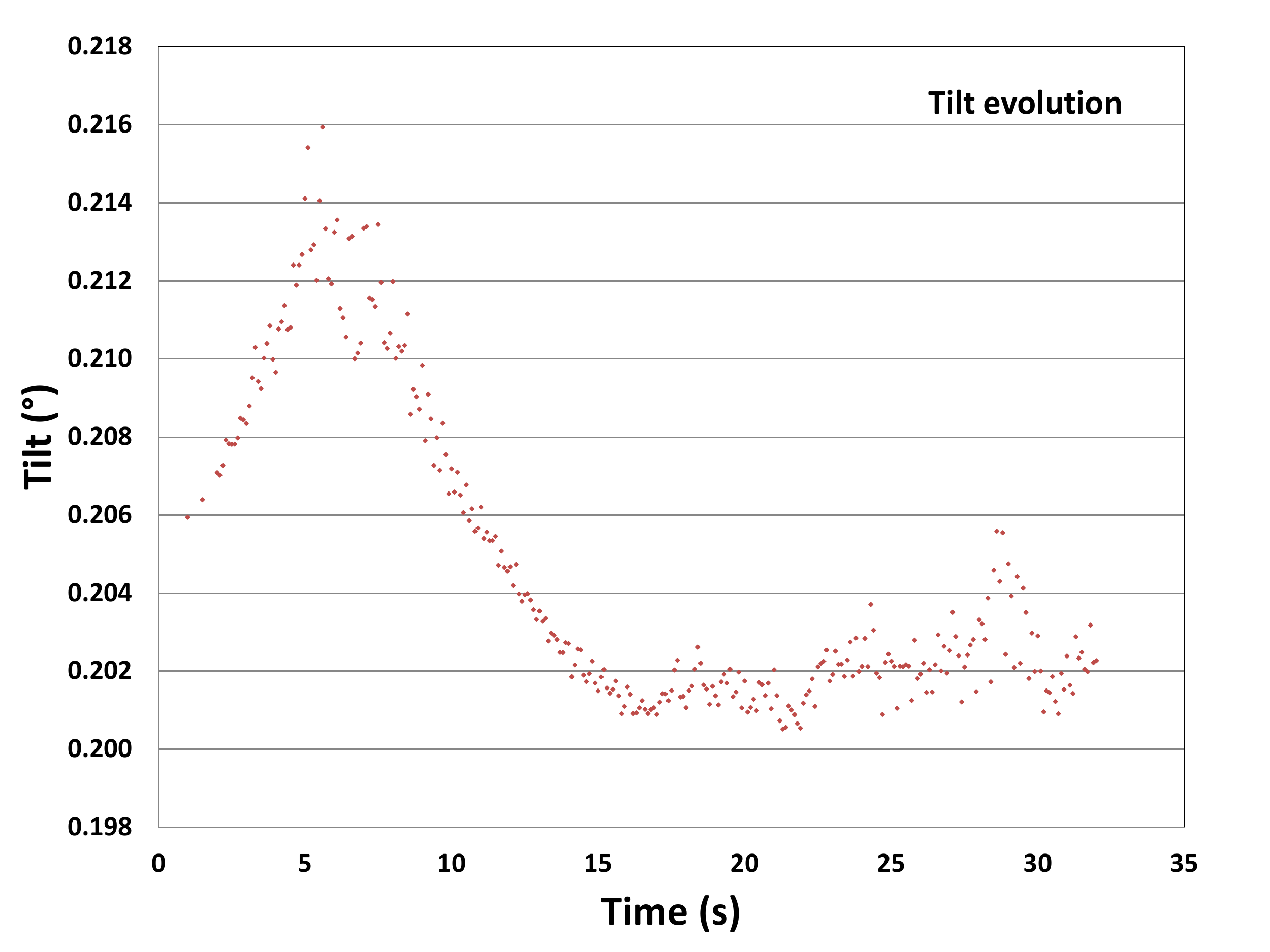}
  \caption{GRB alert azimuth 180$^{\circ}$ rotation plus turbulent wind: focal distance variation, radial displacement and tilt.}
  \label{resu13}
 \end{figure}

A first critical load case combines a Gamma Ray Burst repositioning (e.g. by azimuth movement of 180$^{\circ}$) and turbulent wind, coming from a constant direction transverse to the arch plane. Figure \ref{ino13} shows the angular azimuth acceleration during the 20 s after notification of a GRB alert. This load case has been carried out in order to check several specifications:

a) The structure is able to undergo an azimuth acceleration (corresponding to 180$^{\circ}$ in 20 s) without any deterioration.

b) The addition of the wind loading does not cause significant problems.

c) Ten seconds after the end of the movement, the telescope is able to take data.

According to non-linear algorithm resolution, the first 2 s are used to implement cable tension and gravity (no dynamic load-case). Angular acceleration is then imposed between 2 s and 22 s (according to the requirements). Wind loading is imposed from 2 s until the end (32 s), in order to verify the displacements of the camera 10 s after the end of the movement.

The results of this study are shown in figure \ref{resu13}:

- The three parameters (namely radial displacement, focal distance variation and tilt) have the
same value at the beginning and the end, due to the fact that telescope altitude does not
change. Gravity always acts in the same direction on the structure.

- The static component of the radial displacement is about 20 mm for this elevation. Dynamic
amplitudes of the radial displacement are roughly 5 mm peak to peak.

- The tilt is almost constant despite the angular acceleration, with an allowable value of 0.2$^{\circ}$

It has to be noticed that tensions in all ropes are always strictly positive during the entire duration of the simulated process. The stress level in the different plies of the structure is allowable, taking into account a safety factor.
 
In conclusion, with the current CSS design the LST is able to perform observations and take data immediately after the repositioning procedure. During the entire process and also during the subsequent 10 s the residual displacements of the camera are well below the maximum specifications.

A second load case gamma ray burst repositioning has been carried out in order to check the same specifications as before, but combining a different angular acceleration (90$^{\circ}$ elevation movement) and turbulent wind, as in the previous case. 

\begin{figure}[!]
  \centering
  \includegraphics[width=0.4\textwidth]{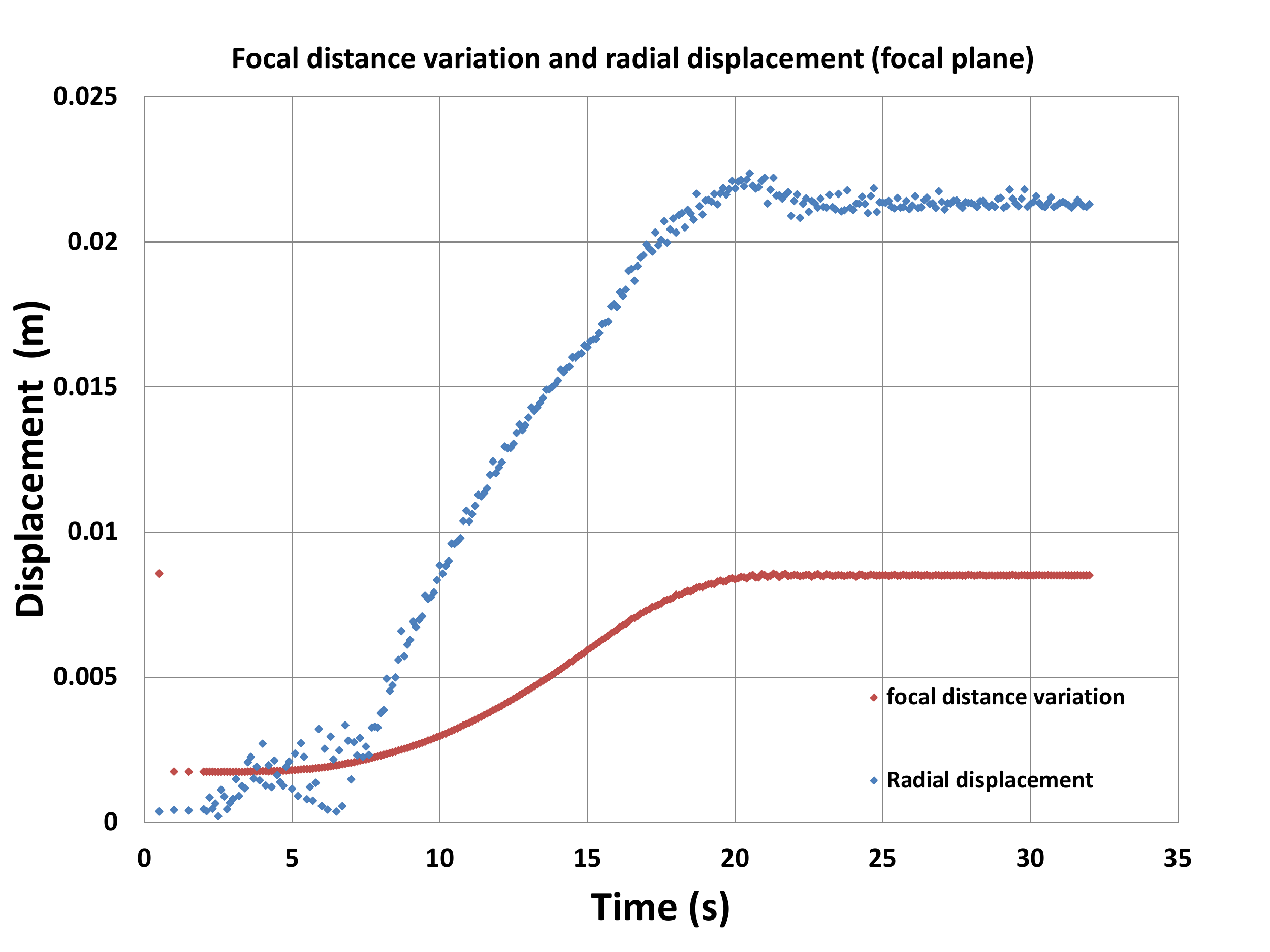}
  \includegraphics[width=0.4\textwidth]{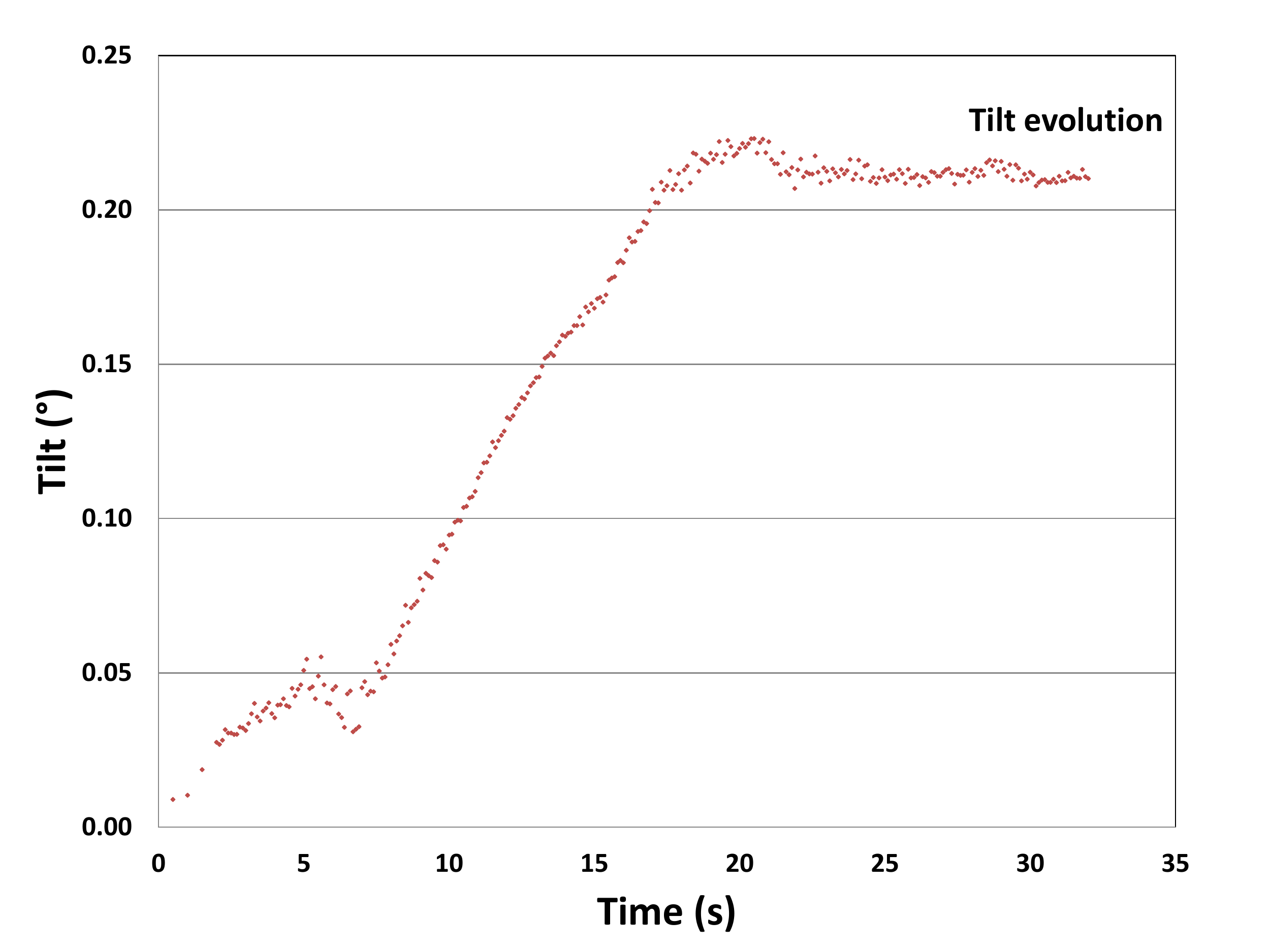}
  \caption{GRB alert zenith 90$^{\circ}$ rotation plus turbulent wind: focal distance variation, radial displacement and tilt.}
  \label{resu15}
 \end{figure}
The dynamic results from this study are shown in figure \ref{resu15}. The radial displacement is negligible before any movement of the telescope (0$^{\circ}$ orientation), and grows when wind and angular acceleration are imposed. Finally its static component
stabilises close to the value of 20 mm, which corresponds to the sagging of the structure at 90$^{\circ}$. This value is in total agreement with the 20 mm obtained for the previous load-case. The dynamic radial displacement oscillates with an amplitude of less than 5 mm at the end of
the computation. Due to the sagging of the structure, the focal distance variation goes from 2 mm to 8 mm when
telescope goes from the zenith to a horizontal direction (0$^{\circ}$ to 90$^{\circ}$). The dynamic effects of the focal distance variation are negligible. Finally, the tilt of the camera follows an increasing slope, with its maximum value when telescope is at 90$^{\circ}$ (and well within the specifications). Tensions in all ropes are always positive in this second case as well.

\section{Conclusions}

The CSS design largely satisfies the specifications and will allow the CTA observatory to follow-up efficiently a GRB alert in a few seconds. 

\vspace*{0.5cm}
\footnotesize{{\bf Acknowledgement:}{This work has been funded by IN2P3, EC-FP7 CTA preparatory phase and it is the result of a collaboration of the authors of this document with $Ryvoire$-$Ing.$
We gratefully acknowledge support from the agencies and organizations listed in this page:http://www.cta-observatory.org/?q=node/22}}


\begin{thebibliography}{}
\bibitem{bib:cta} The CTA observatory, http://www.cta-observatory.org/
\bibitem{bib:oscar} O. Blanch-Bigas et al., ICRC 2013, 776, these proceedings.
\end{thebibliography}
\end{document}